\documentstyle[aas2pp4]{article}

\begin{document}

\title{
Evidence for the Absence of a Stream--Disk Shock Interaction
\linebreak (Hot Spot)
in Semi-Detached Binary Systems. Comparison of Numerical Simulation
Results and Observations}

\author{D.V.Bisikalo\altaffilmark{1}}
\affil{Institute of Astronomy of the Russian Acad. of Sci.,
Moscow, Russia}

\author{A.A.Boyarchuk}
\affil{Institute of Astronomy of the Russian Acad. of Sci.,
Moscow, Russia}

\author{O.A.Kuznetsov\altaffilmark{2}}
\affil{Keldysh Institute of Applied Mathematics, Moscow, Russia}

\author{T.S.Khruzina\altaffilmark{3}}
\affil{Sternberg Astronomical Institute, Moscow, Russia}

\author{A.M.Cherepashchuk}
\affil{Sternberg Astronomical Institute, Moscow, Russia}

\author{V.M.Chechetkin}
\affil{Keldysh Institute of Applied Mathematics, Moscow, Russia}

\altaffiltext{1}{\large E-mail address: {\it bisikalo@inasan.rssi.ru}}
\altaffiltext{2}{\large E-mail address: {\it kuznecov@spp.keldysh.ru}}
\altaffiltext{3}{\large E-mail address: {\it kts@sai.msu.su}}

\begin{abstract}
   The results of three--dimensional numerical simulations of the
flow of matter in non-magnetic semidetached binary systems
are presented. Self-consistent
solutions indicate the absence of a shock interaction between
the stream of matter flowing from the inner Lagrange point and
the accretion disk (a "hot spot"). At the same time, the
interaction between the stream and the common envelope of the
system forms an extended shock wave along the edge of the
stream, whose observational properties are roughly equivalent to
those of a hot spot in the disk. Comparison of synethesized and
observed light curves confirm the adequacy of the proposed model
for the flow of matter without formation of a hot spot. The
model makes it possible for the first time to reproduce the
variety of the observed light curves of cataclysmic variables in
the framework of a single model, without invoking additional
assumptions.
\end{abstract}

\section{Introduction}

   In semidetached close binary systems, in which one component
fills its Roche lobe [1, 2], we can observe the effects of the
flow of matter between the components [3--6], which leads to the
formation of gaseous flows, streams, disks, a common envelope,
and other structures. Especially clear evidence for flows of
matter are observed in semidetached close binary systems in late
stages of their evolution, following the initial fast mass
exchange [7]. For example, we observe very complex light curves
in cataclysmic binary systems consisting of a red dwarf that
fills its Roche lobe and feeds matter through the inner Lagrange
point and a white dwarf surrounded by an optically bright
accretion disk. It is not possible to adequately interpret these
light curves using only simple assumptions about the structure
of the flow of matter in these systems.

\figurenum{1a}
\begin{figure}[h]
\plotone{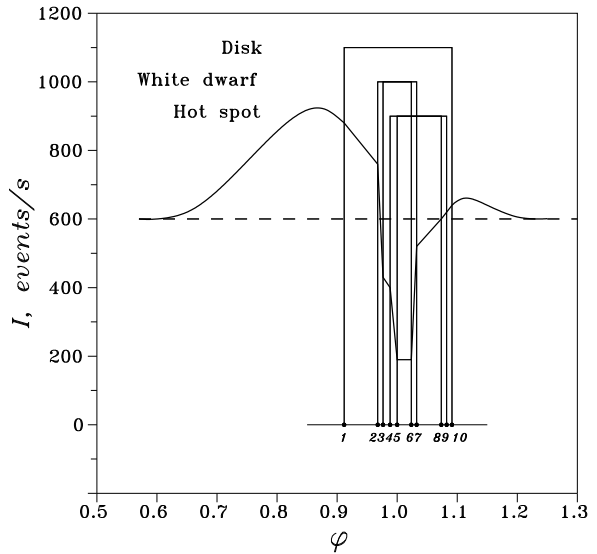}
\caption{Schematic diagram of a typical light curve of a
cataclysmic close binary system with a double eclipse [10].
Phase $\varphi = 0$ corresponds to the time of alignment of the
system components (white and red dwarfs).  The segment 1--2
corresponds to the ingress of the eclipse of the unperturbed
accretion disk, 2--3 to the ingress of the eclipse of the white
dwarf, 3--4 to the continuation of the eclipse of the accretion
disk, 4--5 to a continuation of 3--4 plus the ingress of the
eclipse of the hot spot, 5--6 to the total eclipse of the white
dwarf and hot spot, 6--7 to the egress of the eclipse of the
central region of the accretion disk and white dwarf, 7--8 to
the continuation of the egress of the eclipse of the unperturbed
accretion disk, and 8--9 to the egress of the eclipse of the hot
spot.}
\end{figure}

\figurenum{1b}
\begin{figure}[h]
\plotone{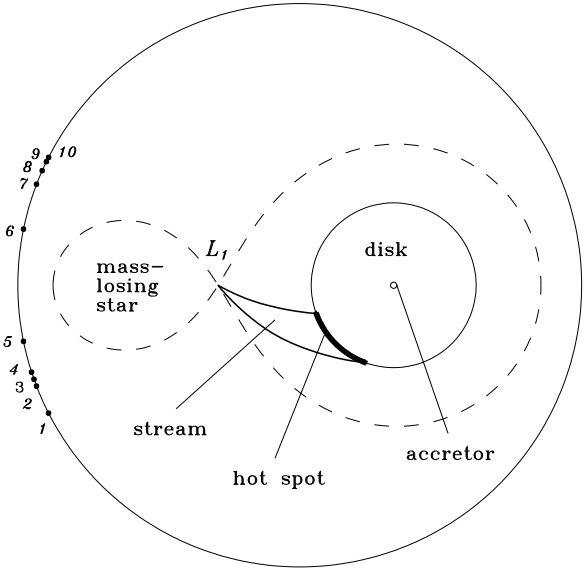}
\caption{Schematic diagram of the main elements of a model of a
semidetached close binary system with a hot spot. The figures
marked on the outer circle correspond to the phase angles
indicated in Fig.~1a for the beginning and end of eclipse for
various elements of the system.}
\end{figure}

   Currently, the best source of information about the flow
structure in binary systems is provided by the light curves of
cataclysmic variables. Humps are observed in the orbital light
curves for cataclysmic close binary systems in their quiescent
state (see, for example, [8]). These humps repeat in a regular
fashion with the orbital period, and have amplitudes reaching a
stellar magnitude or more [9]. In five cataclysmic binary
systems in their quiescent states (Z Cha, OY Car, V2051 Oph, HT
Cas, and IP Peg), a deep so-called "double eclipse" is observed
together with the orbital hump (i.e., eclipse of the central
white dwarf and a hot region at the outer boundary of the
accretion disk by the red dwarf). Figure 1a presents a schematic
diagram of a typical light curve for a ca\-ta\-clysmic close binary
system with a "double eclipse" [10].

   The existence of a so-called "hot spot" is widely
hypothesized to explain the observed light curves. This hot spot
is thought to form during the shock interaction of the matter in
the gaseous stream flowing from $L_1$ with the outer boundary of
the accretion disk (see, for example, [6, 8, 11]). Figure 1b
gives a schematic illustration of the main elements in this
model. In order to make clear the
correspondance between the adopted flow model and the light
curve, the figures in Fig.~1a and 1b indicate phase angles
corresponding to the times of the beginning and end of eclipse
of various elements of the system (the disk, white dwarf, and
hot spot). It is clear from Fig.~1 that a flow model with a hot
spot can qualitatively explain the typical observed light curve
rather well. Note also that interpretations of optical light
curves of the sort shown in Fig.~1a using a hot-spot model can,
in many cases, provide a satisfactory agreement between the
model and observations, and yield reasonable model
characteristics, including the size, luminosity, and position
angle for the hot spot (see, for example, [8, 12]). Of the
numerous papers that have applied the hot-spot hypothesis to
interpretations of observations, we especially note [12--15].

   Over many years, the presence of the hump in the light curve
of cataclysmic variables has come to be considered proof of the
existence of a hot spot on the accretion disk. At the same time,
as observational material has been accumulated, it has become
clear that the standard model is not able to explain many
effects. First of all, the humps in the light curves of
different stars occur at different phases. For example, for most
cataclysmic variables, the eclipse is on the descending branch
of the hump, but there are several systems with eclipses on the
ascending branch of the hump (for example, RW Tri and UX UMa [8,
16]). In addition, in a number of cases (the system OY Car, for
example [17]), two humps are observed in the light curve rather
than only one.  It is not possible to explain the observed
shifts of the phase of the hump in different systems using the
standard model for cataclysmic variables in their quiescent
state. The location of the hot spot (and, accordingly, of the
light curve hump) is determined by the kinematic deviation of
the gaseous stream from the line joining the centers of the
binary system components under the action of the Coriolis force,
and should remain virtually constant [18].

   Spectroscopic observations and Doppler tomography of the
gaseous flows and the accretion disk in cataclysmic close binary
systems (see the reviews [19, 20]) have revealed in a number of
cases the existence of S--wave in the emission line radial
velocity curves, and the presence of a bright, compact region at
the outer boundary of the disk. According to the classical
concept of a hot spot, it should be located in the third
quadrant in the velocity plane. In many
cases, however, Doppler tomography either does not show the
effect of a hot spot at all, or indicates that the spot is not
in the third quadrant, i.e., not in the place where we would
expect a collision between the gaseous stream and the outer
boundary of the disk.

   These are weighty arguments against the hot-spot hypothesis
that provide convincing proof of its inadequacy. In addition, we
note that it is difficult to understand the development of a
shock interaction between the stream of matter and the accretion
disk from a gas--dynamical point of view. Even if the stream of
matter from $L_1$ initially struck a previously existent disk,
with time there would be a reconstruction of the flow morphology
making the interaction between the stream and disk become
shock-free, since the stream is the only source of matter for
the disk. Unfortunately, until recently, there were no detailed
gas--dynamical investigations of self-consistent matter flow
patterns in close binary systems, beginning from the time when
the stream of matter originates to the time of formation of the
disk. The most effective attempts to study the overall flow
pattern in semidetached systems were undertaken in [21--23],
which present three--dimensional numerical simulations over
rather long time intervals. A number of interesting results were
obtained, however the use of the Smoothed Particle Hydrodynamics
method to solve the system of gas--dynamical equations made it
impossible to examine the influence of the common envelope on
the flow pattern: the limitations of this method make it
difficult to investigate flows with large density gradients,
and, consequently, the action of the rarified envelope gas on
the gas--dynamical transfer of matter was not taken into account
entirely correctly.

   In order to investigate the structure of the flows in close
binary systems more correctly, it is necessary to consider the
gas dynamics of the gaseous streams in the framework of a
self-consistent model for the flows in close binary systems. The
first such studies were recently presented by us in [24, 25].
The Total Variation Diminishing method used in [24, 25] to solve
the system of gas dynamical equations is free of the
shortcomings of the Smoothed Particle Hydrodynamics method, so
that it was possible to investigate the morphology of the
gaseous streams in the system and study the influence of the
system's common envelope, in spite of the presence of
substantial density gradients. This three--dimensional modeling
of the gas--dynamical transfer of matter indicated that in a
steady-state, self-consistent model for the flow of matter,
the stream flowing through the vicinity of the inner Lagrange
point smoothly bypasses the gaseous disk that forms around the
accretor, so that there is no shock interaction between the disk
and the stream (i.e., there is no "hot spot"). In addition, we
discovered that the interaction of the common envelope of the
system and the stream leads to the formation of an extended
shock wave along the edge of the stream, whose observational
manifestation may be equivalent to that of a hot spot on the
accretion disk. The results in [24, 25] were obtained for a
low-mass X-ray binary system ($M_1 =0.28 M_\odot$, $M_2= 1.4
M_\odot$, $P_{orb}=1^d.78$, distance between component centers
$a = 7.35 R_\odot$). To determine how universal our flow model is, we
investigate here the flow structure in a cataclysmic binary with
very different parameters ($M_{1} =0.19M_\odot$,
$M_{2}=0.94M_\odot$, $P_{orb}=0^d.074$, $a = 0.78R_\odot$).
The results we present below provide convincing evidence that
the flows in these systems are qualitatively similar to those in
low-mass X-ray binaries. This, in turn, suggests that this type
of flow structure is universal in non-magnetic semidetached
binary systems. Thus, a correct three--dimensional,
gas--dynamical analysis indicates that the standard hot spot is
absent in the steady-state solutions for semidetached close
binary systems (!).

   In summary, we note that the overall flow pattern in close
binary systems in our approach has a very different appearance
than in standard models.  Consequently, if we are able to show
that our model is well-based both from the point of view of
gas--dynamics and the point of view of describing the
observations, we expect not merely a transition from one model
to another, but fundamental changes in our understanding of the
physical processes in close binary systems.

\section{Gasdynamics of flows in semidetached binary systems}

   In [24, 25], we presented results of numerical simulations of
flows of matter in non-magnetic semidetached binary systems,
using a low-mass binary system similar to X1822-371 as an
example. This three--dimensional modeling of the gas--dynamical
transfer of matter allowed us to investigate the morphology of
the gaseous flows in the system, and to study the influence of
the common envelope of the system. We showed that, in a
stationary flow regime, the presence of a common envelope leads
to the absence of a shock interaction of the stream flowing from
$L_1$ and the accretion disk. The stream is deflected by the
common envelope gas, and approaches the disk along a tangent, so
that it does not give rise to a shock perturbation at the edge
of the disk. Since the flow pattern obtained in [24, 25] is
fundamentally different from standard models (which is
especially important in the interpretation of observations), the
question arises of how universal this flow structure is. Here,
we consider the flow pattern in a semidetached binary system
with parameters that are typical for cataclysmic binaries, and
are similar to those for the system Z Cha.

   We have already presented a detailed description of the
numerical model used in [24, 25]. Below, we briefly summarize
the main assumptions of the model used here.

   (1) We consider a cataclysmic binary system similar to Z Cha
with the following parameters [26]: mass of the mass-losing
component (red dwarf) $M_{rd}= 0.19M_\odot$; gas temperature at the
red dwarf surface $T=5 \cdot 10^3$~K; mass of the secondary (white
dwarf), which has radius $0.009R_\odot$,
$M_{wd}=0.94M_\odot$; orbital period of the system
$P_{orb}=0^d.074$; and distance between the component centers
$a = 0.776R_\odot$.

 (2) We assumed that the magnetic field was small and does not
exert a significant influence on the flow of matter in the
system.

 (3) We used a three--dimensional system of gas--dynamical
equations to describe the flow, supplemented by the equation of
state for an ideal gas.

 (4) When allowing for the radiative losses in the system, the
adiabatic index was taken to be close to unity ($\gamma=1.01$),
which is close to the isothermal case [27].

 (5) We assumed that the mass-losing star fills its Roche
lobe, and that the gas velocity at its surface is equal to the
local sound speed. The density $\rho$ at the surface of this
component was denoted $\rho_0$. The boundary value of the
density did not affect the solution, due to the scaling of the
system of equations with $\rho$ and $P$. In the calculations, we
used an arbitrary value for $\rho_0$; when considering the real
densities in a specific system with known mass loss rate, the
calculated density values must be scaled in accordance with the
real and model densities at the surface of the mass-losing
component.

 (6) We adopted free outflow conditions at the accretor and the
outer boundary of the calculation region.

 (7) The calculation region was the parallelopiped
$(-a..2a)\times(-a..a)\times(0..a)$; due to the symmetry of the
problem relative to the equatorial plane, we carried out the
calculations only in the upper half-space.

 (8) We used a high-order total variation diminishing scheme to
solve the system of equations on an non-uniform (finer along
the line joining the centers of the system components)
difference grid with $84\times 65\times 33$ nodes.

 (9) We followed the solution from arbitrary initial conditions
right up to the establishment of a steady-state flow regime.
Note that the characteristic gas--dynamical time for
establishment of the flow (the ratio of the characteristic size
of the system $a$ to the velocity of propagation of
perturbations, which is of the order of the local sound speed at
the surface of the mass-losing component) is of the order of
nine orbital periods. We therefore carried out the calculations
over a substantially longer time interval, more than 15 orbital
periods, in order to be certain that we had attained a
stationary solution. We verified the stationarity of the
solution using both local and integrated characteristics of the
flow.

\figurenum{2}
\begin{figure}[h]
\plotone{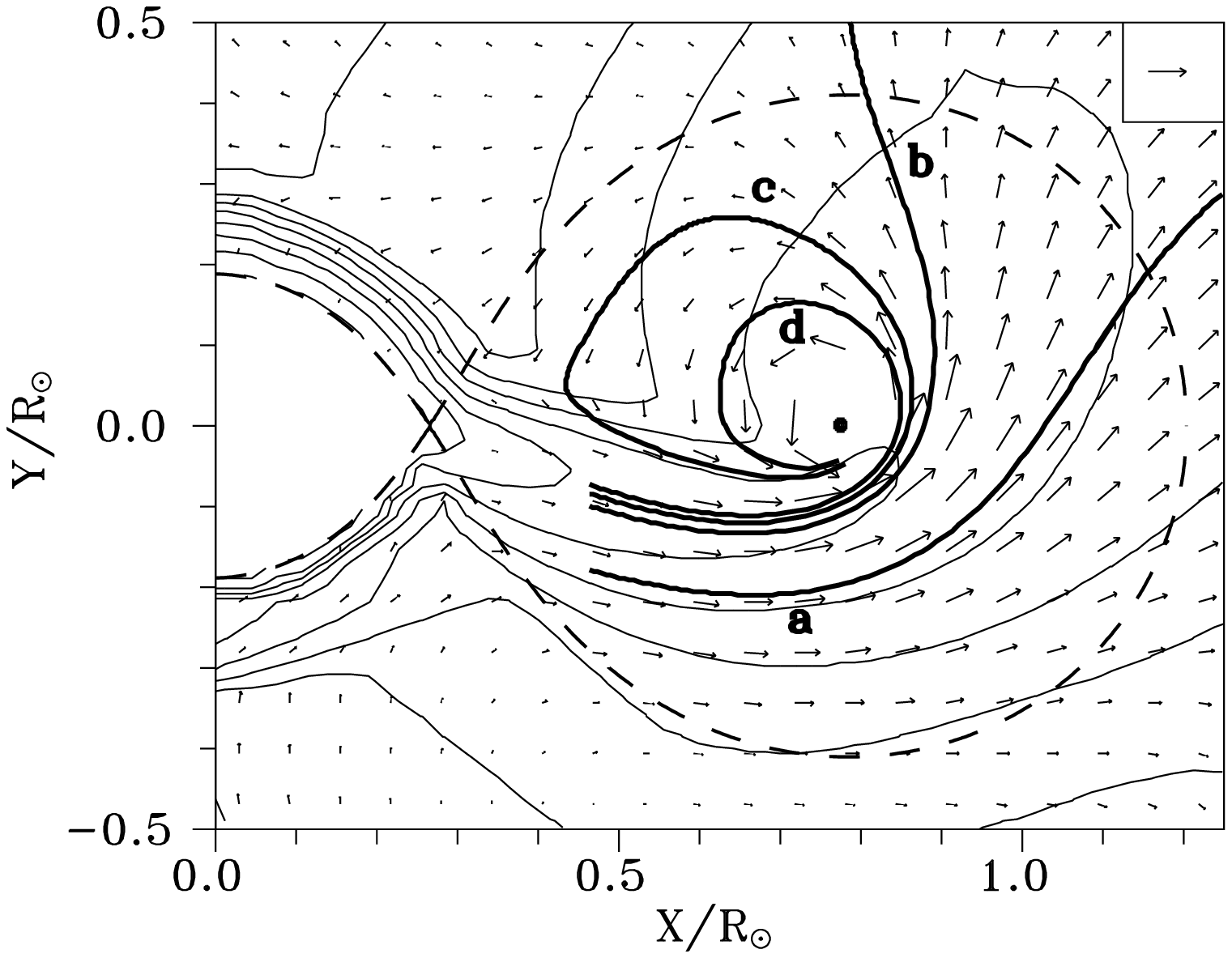}
\caption{Density isolines and velocity vectors in the equatorial
plane of the system. Roche equipotentials are shown with dashed
lines. Four stream lines marked with the letters {\bf a}, {\bf b}, {\bf
c}, {\bf d} are also shown, illustrating directions of the flow of the
common envelope gas. The location of the accretor is marked with the
dark circle. The vector in the upper right corner corresponds to a
velocity of 1200 km/s.}
\end{figure}

   The overall flow pattern in the equatorial plane of the
system is presented in Fig.~2, which depicts density contours
and the velocity field in the region from 0 to $1.25R_\odot$
along the $X$ axis and from $-0.5$ to $+0.5R_\odot$ along the
$Y$  axis.  Figure 2 also shows four flowlines marked by the
symbols {\bf a}, {\bf b}, {\bf c}, {\bf d}, which illustrate the
directions of flows in the system.  The results presented in Fig.~2
enable us to conditionally divide the matter in the flows into three
parts.  The first of these (flowline {\bf d}) forms a quasi-elliptical
accretion disk, and further loses angular momentum to viscosity and
participates in the accretion process.  The second (flowlines {\bf b}
and {\bf c}) passes around the accretor beyond the disk. The third
(flowline {\bf a}) moves in the direction of the Lagrange point $L_2$;
a significant fraction of this matter changes the direction of its
motion under the action of the Coriolis force and remains in the
system.

   We can estimate the linear size of the disk from the last
flowline along which matter falls directly onto the disk. Along
the previous flowline (between {\bf c} and
{\bf d}), matter passes around
the accretor and interacts with the stream, though it may then
(after interacting with the stream) also be accreted. The
marginal flowline in Fig.~2 is {\bf d}, and it is not difficult
to determine the size of the resulting quasi-elliptical disk,
which proves to be $0.25 \times 0.22 R_\odot$ (or $0.32 \times
0.28 a$).  Analysis of the numerical modeling indicates that the
thickness of the disk varies from 0.006 to $0.04 R_\odot$ (or
from 0.7 to 4.5 times the radius of the accretor).

   The most influence on the overall flow pattern is exerted by
that part of the matter that remains in the system but is not
directly involved in the accretion process (flowlines {\bf a}, {\bf b},
and {\bf c} in Fig.~2). In accordance with the terminology in our
previous papers [24, 25], we will call this material the common
envelope of the system. Note that a substantial fraction of the
common envelope gas (flowlines {\bf a} and {\bf b}) interacts with the
matter flowing from the surface of the donor star. The affect of
this part of the common envelope on the flow structure
significantly changes the mass exchange regime in the system. A
detailed discussion of this effect is presented in [25].

   The remainder of the common envelope (flowline {\bf c}) passes
around the accretor and undergoes a shock interaction with the
edge of the stream that faces into the orbital motion. This
interaction also leads to significant changes in the overall
flow pattern in the system. Analysis of the variations in the
parameters for the gas flowing along the flowlines shown in
Fig.~2 indicates that the flow is smooth for all the flowlines
that belong to the disk, up to the boundary flowline {\bf d}. The
absence of breaks in the smooth flow provides evidence that the
interaction of the stream and the disk is shock-free, which, in
turn, implies the absence of a hot spot at the edge of the disk.
The origin for this shock-free morphology for the stream--disk
system can be seen in Fig.~2, which clearly shows that the
stream of matter is deflected by the common envelope gas
(flowline {\bf c}) and approaches the disk tangentially. At the same
time, as noted above, the interaction of the stream and the
common envelope forms an extended shock front along the edge of
the stream facing into the orbital motion. The parameters of
this shock and the total amount of energy released in it can be
estimated from the calculation results.

\figurenum{3}
\begin{figure}[h]
\plotone{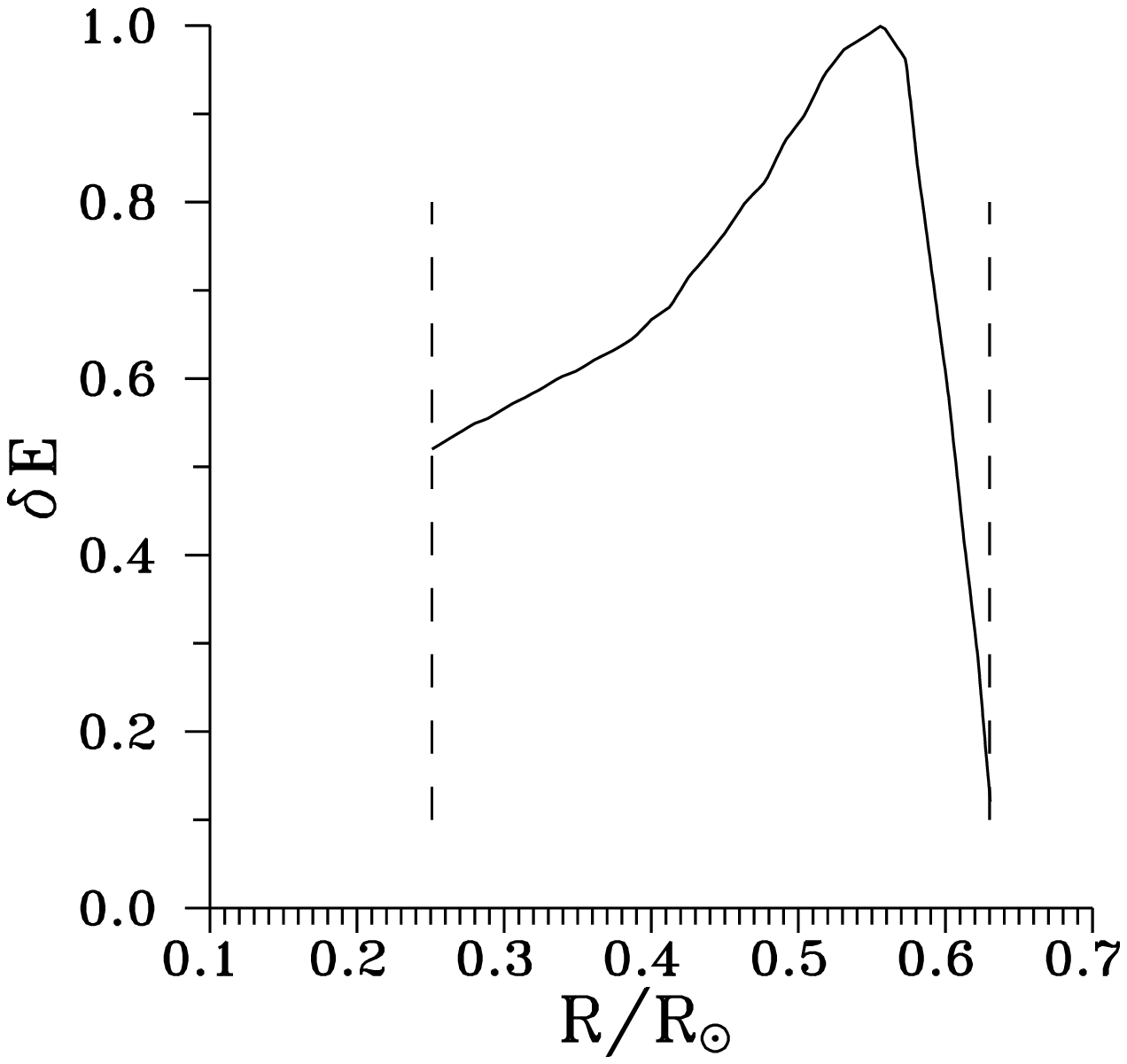}
\caption{
Distribution of the specific rate of energy release $\delta E$
along the shock wave in the equatorial plane, normalized to
unity. The dashed lines mark the shock boundaries.}
\end{figure}

\figurenum{4}
\begin{figure}[h]
\plotone{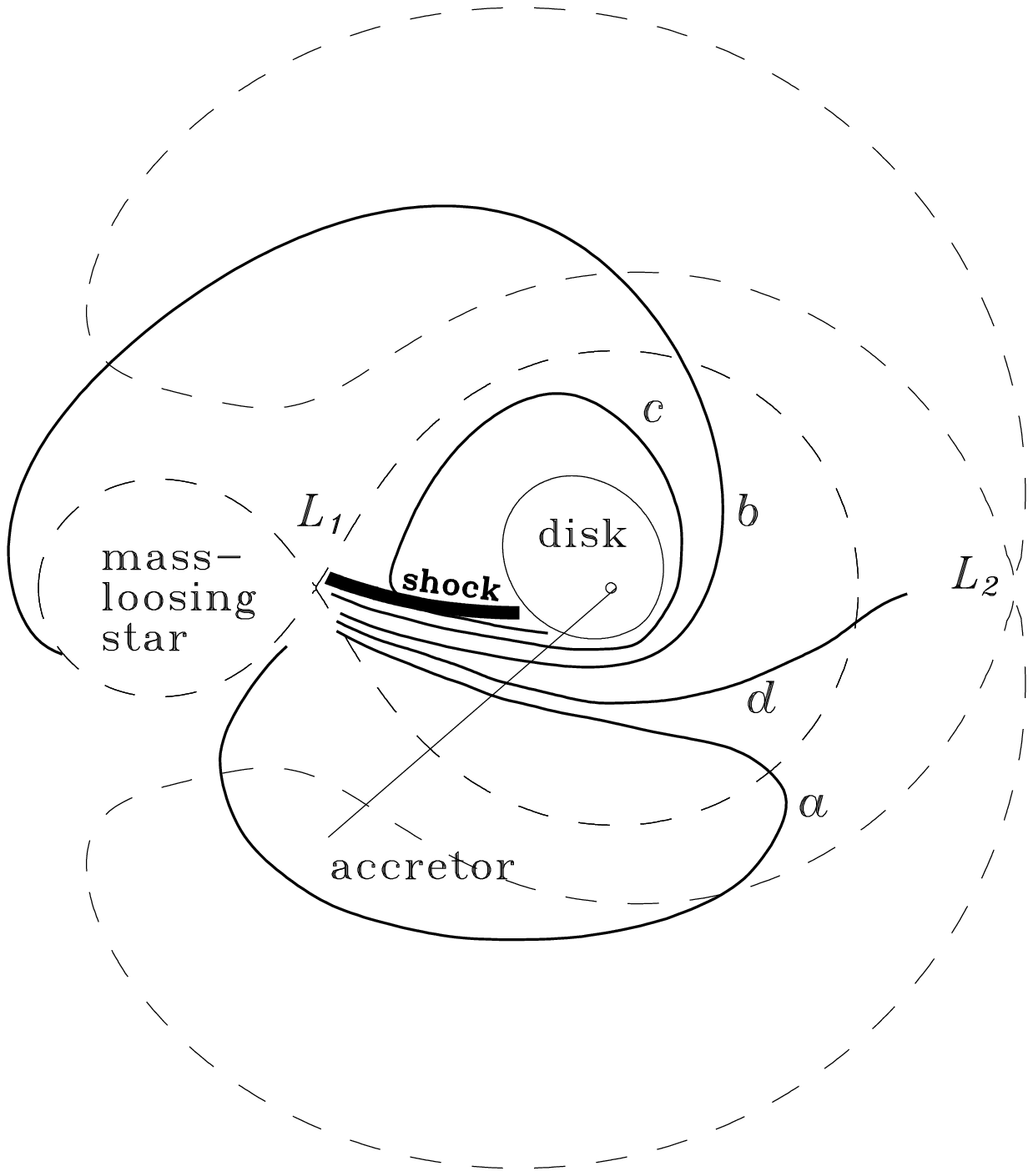}
\caption{Schematic diagram of the main properties of the gas--dynamical
flow pattern in semidetached close binary systems. The Roche lobe
(dashed curve), accretor, Lagrange point, and quasi-elliptical
accretion disk are shown. The shock wave that arises during the
interaction of the common-envelope gas with the stream is shown
by the thick line. The flowlines {\it a}, {\it b}, {\it c}, and
{\it d} illustrate the main directions of the gas flows in the
system.  The flows {\it a}, {\it b}, and {\it c} form the common
envelope of the system, while material flowing along flowline
{\it d} leaves the system.}
\end{figure}

   One of the characteristic features of the shock is the
variation of its intensity along the stream. This is illustrated
in Fig.~3, which presents the distribution of the specific
energy release rate $\delta E$ (erg$\cdot$s$^{-1}\cdot$cm$^{-2}$)
along the shock wave in the equatorial plane, normalized to
unity. The vertical dashed lines in Fig.~3 show the boundary of
the stream:  the line to the left shows the point at which the
stream begins --- the Lagrange point $L_1$, while the line to
the right shows the ending point of the stream, i.e., the place
at which the stream and disk come into contact. We can see from
Fig.~3 that the main energy release in the system occurs in a
compact region of the shock wave near the accretion disk. The
total rate of energy release in the shock $\Delta E_{shock}$ is
comparable with estimates of the energy release in a standard
hot spot $\Delta E_{spot}$, calculated assuming the hot spot is
at the place of contact between the stream and disk.

   The main features of the calculated flow pattern for a
cataclysmic system similar to Z Cha coincide with those for the
low-mass binary system presented in [24, 25]. This suggests that
this type of flow structure is universal for stationary
non-magnetic semidetached binary systems. The main
properties of this flow structure are summarized in the
schematic diagram shown in Fig.~4.

   Note that these results were obtained for a steady-state flow
regime. In a non-stationary regime, when the flow morphology is
determined by external factors or by an unusual distribution of
the parameters characterizing the viscosity in the disk and is
not self-consistent, regions of shock interaction between the
disk and gas flows in the system may appear. For example, if the
disk is formed before the donor star fills its Roche lobe, a hot
spot can arise at the place of contact between the stream and
the outer edge of the disk after the initial stage of mass
transfer through the vicinity of $L_1$.  Since we expect a
self-consistent solution without a hot spot after the flow has
entered the steady-state regime, the principle issue is the life
time of this formation. A natural characteristic life time for
the hot spot is the interval over which the quantity of matter
carried into the system by the stream becomes comparable to the
mass of the accretion disk, since after the transfer of the disk
material, the solution becomes self-consistent. Our estimates in
[24] indicated that for mass--transfer and accretion disk
parameters typical for semidetached binary systems, we expect
that the time for the system to reach a steady-state regime is
not large (of the order of tens of orbital periods).
Consequently, the probability of observing a hot spot is small,
and the observations are dominated by the energy released in the
shock at the edge of the stream.

   Thus, in place of earlier standard models for close binary
systems that assumed the presence of a hot spot at the outer
boundary of the disk, we propose a model in which the region of
energy release is outside the accretion disk. This region arises
as a result of a shock interaction between the common envelope
gas passing around the accretor and the stream. To test the
adequacy of this model, we synthesized light curves and compared
them with observations. The details of this comparison are
presented in the following section.

\section{Interpretation of Light Curves of Cataclysmic Binaries}

   It is clear that light curves currently represent the most
complete and informative data for cataclysmic binaries.
Calculating realistic light curves for these complex stellar
systems, which include two stars, a common envelope, an
accretion disk, gaseous flows, and shock fronts, is a very
labor--intensive task. To ensure that these calculations are as
correct as possible, we must consider the transfer of radiation
in an inhomogeneous medium with large gradients of the gas
parameters (temperature, density, and velocity) and optical
depth. Here, we set ourselves a more modest task --- to
demonstrate by varying certain parameters of the system that it
is possible to qualitatively explain the variety of the light
curves of cataclysmic close binary systems in the framework of
our gas--dynamical model for the flow of matter in these
systems.

\subsection{Description of the Photometric Model Used}

   In our photometric calculations, we assumed that the system
consisted of two stars (a normal star that fills its Roche lobe
and a white dwarf), an accretion disk, and a bulge on the disk,
which approximates the region of emitting gas in the shock
front. To simplify the calculations, we assumed that the shape
of the bulge is described by part of an ellipsoid of rotation
with semiaxes $a_b$ and $c_b$. The axis $a_b$ lies in the
orbital plane, while the $c_b$ axis --- the rotational axis ---
is perpendicular to the orbital plane. The center of the bulge
is specified by its azimuth, i.e., the angle between the line
joining the system components and the radius vector from the
center of the white dwarf to the center of the bulge
$\alpha_b$.  Figure 5 shows the relative locations of the main
elements of the system in the equatorial plane.

\figurenum{5}
\begin{figure}[h]
\plotone{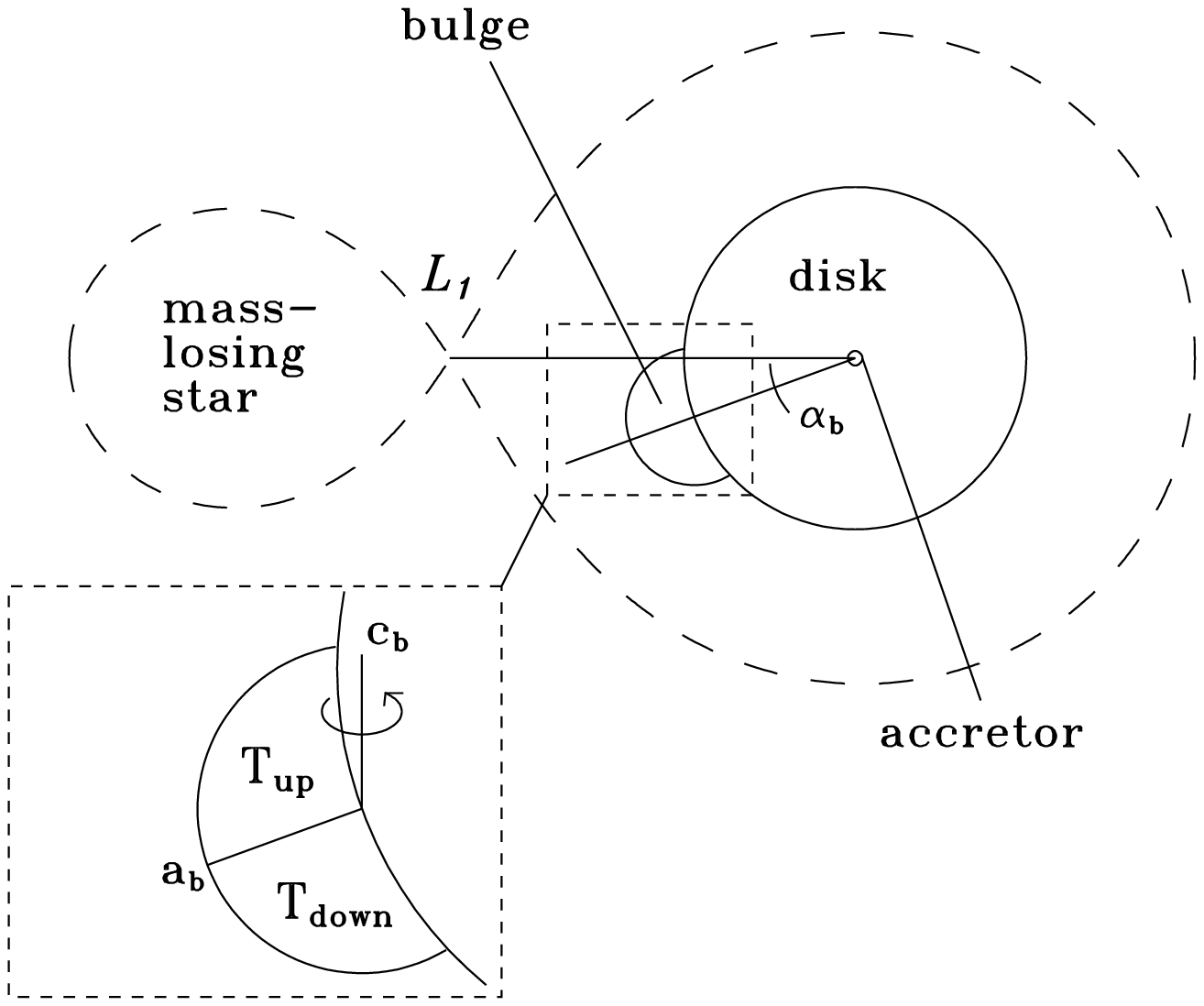}
\caption{Geometrical model of a cataclysmic variable with a bulge at
the edge of the accretion disk. The model is described in the text.}
\end{figure}

   Unfortunately, the temperature distribution over the surface
of the bulge and the optical depth along various directions
cannot be calculated in the framework of our model. Therefore,
to determine the luminosity of the bulge, we divided it into two
halves by the radius vector from the disk center, with one half
above (on the side of the incident gas stream) and the other
below the radius vector. The temperatures of these two halves
($T_{up}$ and $T_{down}$, respectively) could be different. Note
that the case when only the lower part of the bulge emits more
closely corresponds to the location of the region of energy
release in the shock wave indicated by our gas--dynamical model
calculations. However, the use of $T_{up}$ and $T_{down}$ to
arbitrarily vary the luminosity of the upper and lower halves of
the bulge not only models the temperature distribution over the
bulge surface (the location of the region of energy release),
but also models various conditions for the visibility of the
bulge. The shape of the energy release region in our photometric
model does not fully correspond to the shape of the shock wave
at the edge of the gaseous stream, but its location outside the
accretion disk makes it possible to conduct a qualitative
analysis of the adequacy of our gas--dynamical flow model.

   Other parameters in the photometric model were the ratio of
the masses of the two stars $q = M_{wd}/M_{rd}$, the
inclination of the orbit of the system $i$, and the temperatures
of both stars $T_{rd}$ and $T_{wd}$. We specified all sizes and
distances as dimensionless quantities in units of the distance
between the components $a$. We also assumed that the red dwarf
fills its Roche lobe and that the white dwarf can be described
by a sphere of radius $R_{wd}$.  The accretion disk surrounding
the white dwarf had a specified radius $R_d$ and
half-thickness at its outer edge $h_d$. The brightness
temperature of the gas at the inner edge of the disk was one of
the parameters to be determined, and enabled calculation of the
radial dependence of the disk temperature, in particular, of the
temperature of the disk at its outer edge $T_d$. We assumed
that the spectra of both stars, various parts of the disk, and
the bulge could be described by Planck distributions. We also
took into account the mutual heating of the components of the
binary system.

   A similar photometric model was proposed by Khruzina [28],
and employed to calculate the light curves of an X-ray source
[29] and cataclysmic binary system [30]. Note that, in [28--30],
the application of models with an region of energy release
outside the disk was not based on gas--dynamical calculations,
but was chosen purely intuitively, as a way to provide agreement
between the observations and the model light curves. In
particular, comparison of the synthesized light curves for
models with a standard hot spot and with a bulge on the disk
with observed light curves for IP Peg [30] showed that the model
with the region of energy release outside the disk fit the
observed light curve much better than the model with a hot spot
at the edge of the disk.

\subsection{Comparison of the Model and Observed Light Curves}

 The shape of the light curve for each dwarf nova is different;
however, given the repeated presence of a number of
characteristic features, these systems can be divided into
several large groups.

 (1) A large fraction of dwarf novae, whose orbital inclinations
$i$ are close to zero, show neither a hump nor an eclipse in
their light curves.  Irregular brightness variations that do not
show any orbital periodicities are characteristic of these
objects.

\figurenum{6}
\begin{figure}[h]
\plotone{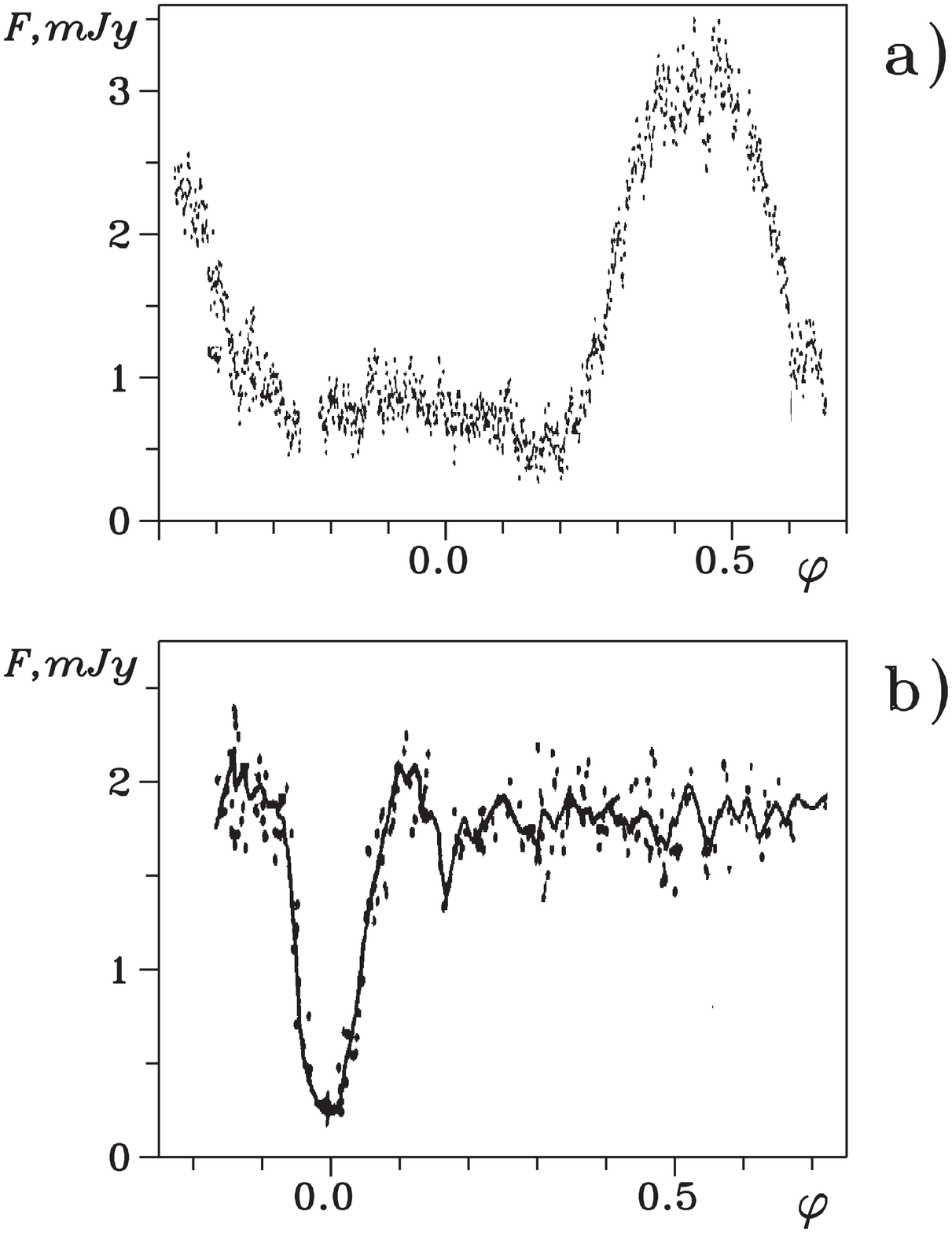}
\caption{Observed light curves for the dwarf novae: (a) VW Hyi,
with the characteristic hump in the light curve, but without an
eclipse [31], and (b) AC Cnc, with an eclipse, but without the
hump [33].} \end{figure}

 (2) Many dwarf novae show an obvious hump in their light curve,
but without any traces of an eclipse. One example of such a
system is VW Hyi [31] (Fig.~6a), which is a SU UMa binary. The
relative and absolute amplitudes of its light curve humps vary
from cycle to cycle, and when the brightness of the system as a
whole increases, the hump amplitude increases as well. If the
hump is present, it usually occupies roughly half the orbital
period, and the beginning and end of the hump occur at about the
same orbital phase.

  (3) In many objects whose orbital inclination $i$ differs
strongly from zero, we observe a rapid drop in brightness in the
light curve --- an eclipse. As a rule, the eclipse occurs soon
after the maximum of the main hump in the light curve (a typical
example of such a system is U Gem [32]). The total duration of
the eclipse is about 0.1 orbital cycles. The exact shape of the
eclipse is usually slightly variable if different minimum for a
single object are compared, but on average maintains its shape.
Note that in some objects, the eclipses are irregular, however
there is no hump in the light curve (for example, the system AC
Cnc; Fig. 6b [33]), and in some eclipsing systems, the hump
appears episodically.

 (4) As a rule, if there is an eclipse, it occurs at the time of
the maximum of the hump or just after it.  However, in some
nova-like stars (UX UMa type stars, similar to BV Cen and OY Car
[17]), an additional intermediate hump is occasionally observed
approximately between the end of the main hump and the beginning
of the following one. This intermediate hump usually has more
moderate brightness than the main hump, and its shape and
amplitude are very variable. The beginning and end of both the
main and the intermediate humps are repeatable with high
accuracy.

 (5) Five dwarf novae whose orbital inclinations are close to
$90^\circ$ (Z Cha, OY Car, V2051 Oph, HT Cas, and IP Peg) have a
very specific eclipse --- a double eclipse. Both the ingress
and egress of the eclipse occur in two stages, with the egress
usually significantly longer than the ingress. Such double
eclipses provide a large amount of information about the system,
and are therefore objects of intense study. The shape of double
eclipses can change from cycle to cycle. Here, it is obvious
that two bright sources are eclipsed:  the first sharp decrease
in brightness is the eclipse of the white dwarf, after which, as
usually assumed, follows the eclipse of the hot spot. After a
more or less flat minimum, we observe the egress of the white
dwarf eclipse followed by the extended egress of the hot-spot
eclipse. The phase of the hot-spot eclipse egress changes from
cycle to cycle. The phase of the primary minimum is highly
stable in all objects.

 In spite of the variety of the shapes of these light curves,
their main features can be qualitatively well reproduced in the
framework of our gas--dynamical model, with an energy release
zone outside the accretion disk.  Figures 7--11 present observed
and theoretical light curves for cataclysmic variables with the
most characteristic properties. We emphasize that we have
limited our treatment to a qualitative comparison of the
synthesized and observed light curves. The parameters of the
binary systems used in our photometric calculations are
summarized in the Table~1.

\begin{table*}
\begin{center}
\begin{tabular}{|c|c|c|c|c|c|}
\hline
     &            &  Z  Cha  & U Gem & RW Tri & OY Car\\
\hline
\multicolumn{6}{|c|}{Common system parameters}\\
\hline
$q$ & Mass ratio   &    4.95       &    2.2      &     1.2
& 9.8\\
\hline $i$ & Orbit incl.  &   $80^\circ$  &  $72^\circ$ &
$74^\circ$  & $83^\circ$\\
\hline
\multicolumn{6}{|c|}{Mass-losing star}\\
\hline
$T_{rd}$& Temperature &    4500~K    &   3100~K    &    3700~K &
3000~K\\
\hline \multicolumn{6}{|c|}{Accretor}\\
\hline
$R_{wd}$& Radius & 0.0158$a$ & 0.0163$a$ & 0.0104$a$ & 0.0107$a$\\
\hline
$T_{wd}$&Temperature&   33000~K     &   28000~K   &    45000~K
& 35000~K\\
\hline
\multicolumn{6}{|c|}{Accretion disk}\\

\hline
$R_d$ & Radius & 0.395$a$ & 0.32$a$ & 0.233$a$ & 0.354$a$\\
\hline
$h_d$ & Thickness &  0.013$a$      &    0.034$a$  &     0.028$a$
& 0.014$a$\\
\hline $T_d$ & Temperature & 2300~K & 3000~K &
3400~K & 2500~K\\

\hline

\multicolumn{6}{|c|}{Bulge}\\

\hline
$a_b$&Large semiaxe& 0.103$a$ & 0.223$a$ & 0.152$a$ & 0.071$a$\\
\hline
$c_b$&Small semiaxe& 0.013$a$ & 0.034$a$ & 0.032$a$ & 0.016$a$\\
\hline
$\alpha_b$& Azimuth & $14^\circ$ & $14^\circ$ & $22^\circ$ &
$25^\circ$\\
\hline $T_{up}$&Temperature & 7100~K & 4300~K &
37000~K & 7500~K\\
 &of upper part     &            &             &
 &           \\
\hline $T_{down}$&Temperature& 21000~K  &
17000~K  &    17000~K    &    12500~K\\
 & of down part    &
      &             &               &           \\
 \hline
\end{tabular}
\end{center}
\caption{The parameters of the
binary systems used in photometric calculations}
\end{table*}

\subsubsection{Systems with a double eclipse}

  The most interesting cases from the point of view of verifying
the applicability of our model to the analysis of close binary
systems are cataclysmic variables with double eclipses. We chose
one such system --- Z Cha --- as a basis for our calculations of
the flow pattern of matter in our gas--dynamical model (Section
2). When constructing the theoretical light curve for this
system, we fixed the ratio of the component masses, orbital
inclination, temperature of the normal star, and radius of the
accretion disk. We fit the parameters of the white dwarf (its
radius and temperature, and the thickness of the outer edge of
the accretion disk) and also the parameters of the bulge in
order to obtain the best qualitative agreement with the observed
light curve (Fig.~7a), which we took from [9].  Figure 7b shows
the theoretical light curve calculated for the parameter values
listed in the Table~1.

\figurenum{7}
\begin{figure}[h]
\plotone{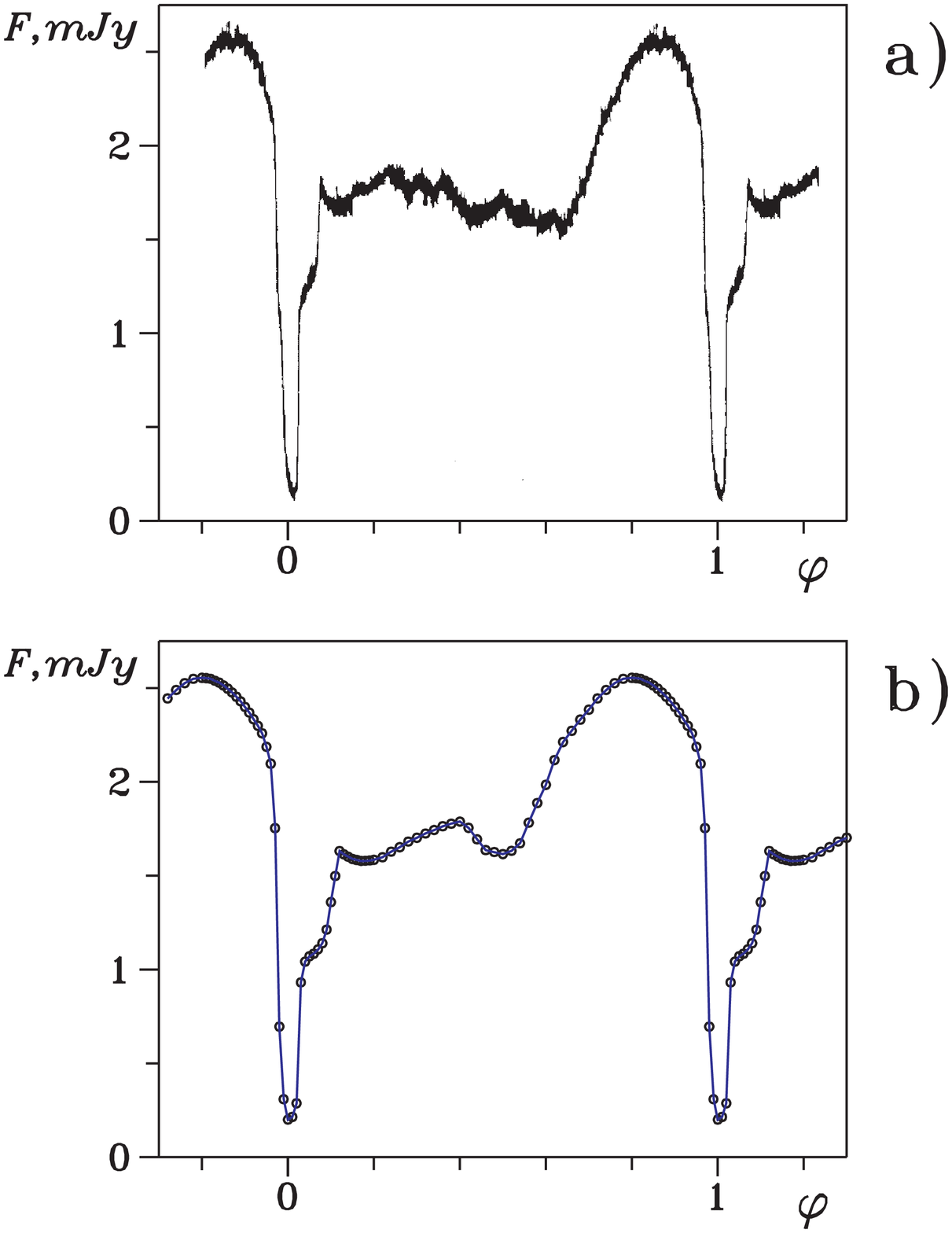}
\caption{Cataclysmic variable with a double eclipse: (a) observed
[9] and (b) theoretical light curves for the dwarf nova Z Cha in the
$V$ filter.}
\end{figure}

   A visual comparison of the curves in Fig.~7 shows that they
are in good qualitative agreement. We can see in the theoretical
curve essentially all the main features in the observed light
curve of Z Cha. The egresses for both the white dwarf and bulge
eclipse are well followed. The modest minimum at phase $\varphi =
0.5$ is associated with the ellipsoidality effect of the red
dwarf. Note that the fitted parameters for the white dwarf
$R_{wd} = 0.0158a$ and $T_{wd} = 33 000$~K are close to the
values derived earlier in [26], $R_{wd} = 0.0116a$ and $T_{wd} =
17 000 \div 24 000$~K.

\subsubsection{Systems with a single hump and ordinary eclipse}

   The shape of the observed light curve in the system U
Gem (Fig.~8a), in contrast to the light curve for Z Cha, does
not show any trace of the eclipse of the white dwarf, and
corresponds to the eclipse only of the region of energy release
(the bulge). Using the parameters obtained earlier for U Gem
($q = 2.2$, $T_{rd} = 3100$~K (SP M4 V), $R_d = 0.32a$ [7]), we
were able to obtain good qualitative agreement between the
observed and theoretical light curves (Figs.~8a, 8b) purely by
increasing the previously proposed inclination value $i =
69^\circ\pm 1^\circ$ to $72^\circ$. The remaining parameters
for the photometric light curve are presented in the Table~1. As
expected, the bulge dominates the total brightness of the
system, and variations in its brightness with orbital phase
determine the appearance of the light curve.  At this orbital
inclination, the white dwarf is not eclipsed, in spite of the
fact that the red dwarf is larger than in the Z Cha system (the
radius of the red dwarf in U Gem is $0.3a$, while the size of
the red dwarf in Z Cha does not exceed $0.23a$).

\figurenum{8}
\begin{figure}[h]
\plotone{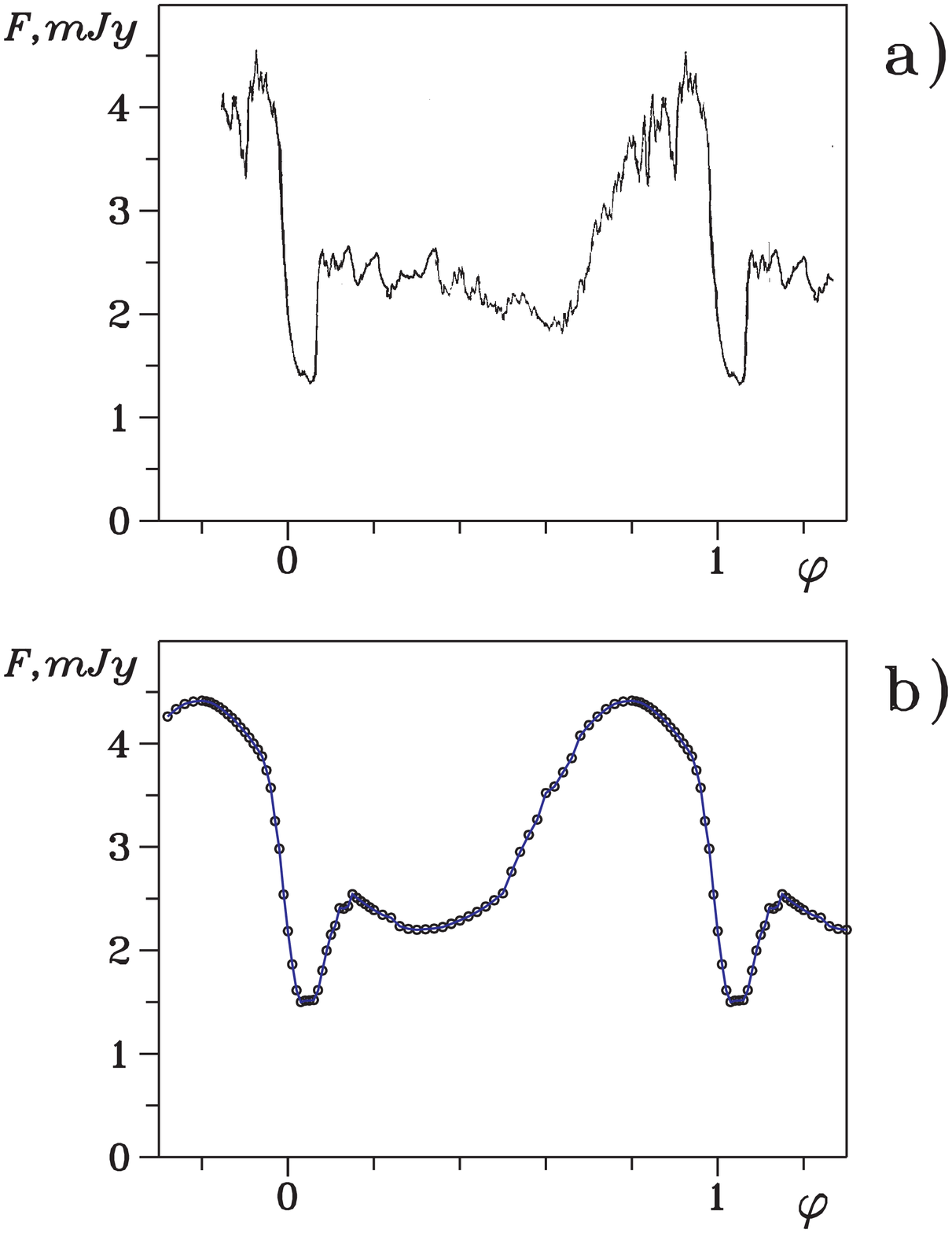}
\caption{System with a single hump and an ordinary eclipse: (a)
observed [32] and (b) theoretical light curves of the dwarf nova U Gem
in the $V$ filter.}
\end{figure}

\figurenum{8c}
\begin{figure}[h]
\plotone{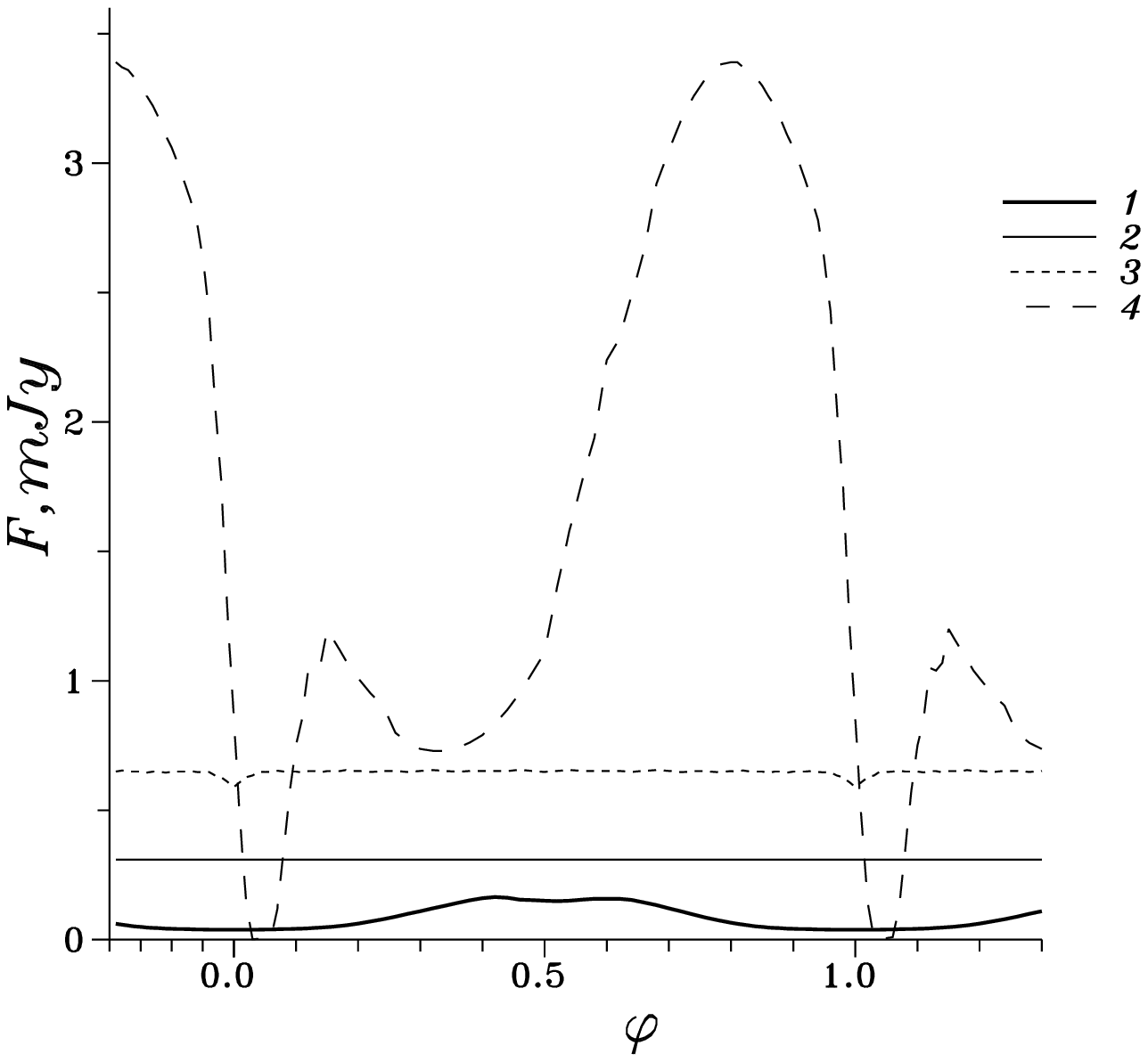}
\caption{Separate dependences of the brightnesses of the red
dwarf (1), white dwarf (2), accretion disk (3), and bulge (4) on
orbital phase for Fig.~8b} \end{figure}

\figurenum{9}
\begin{figure}[h]
\plotone{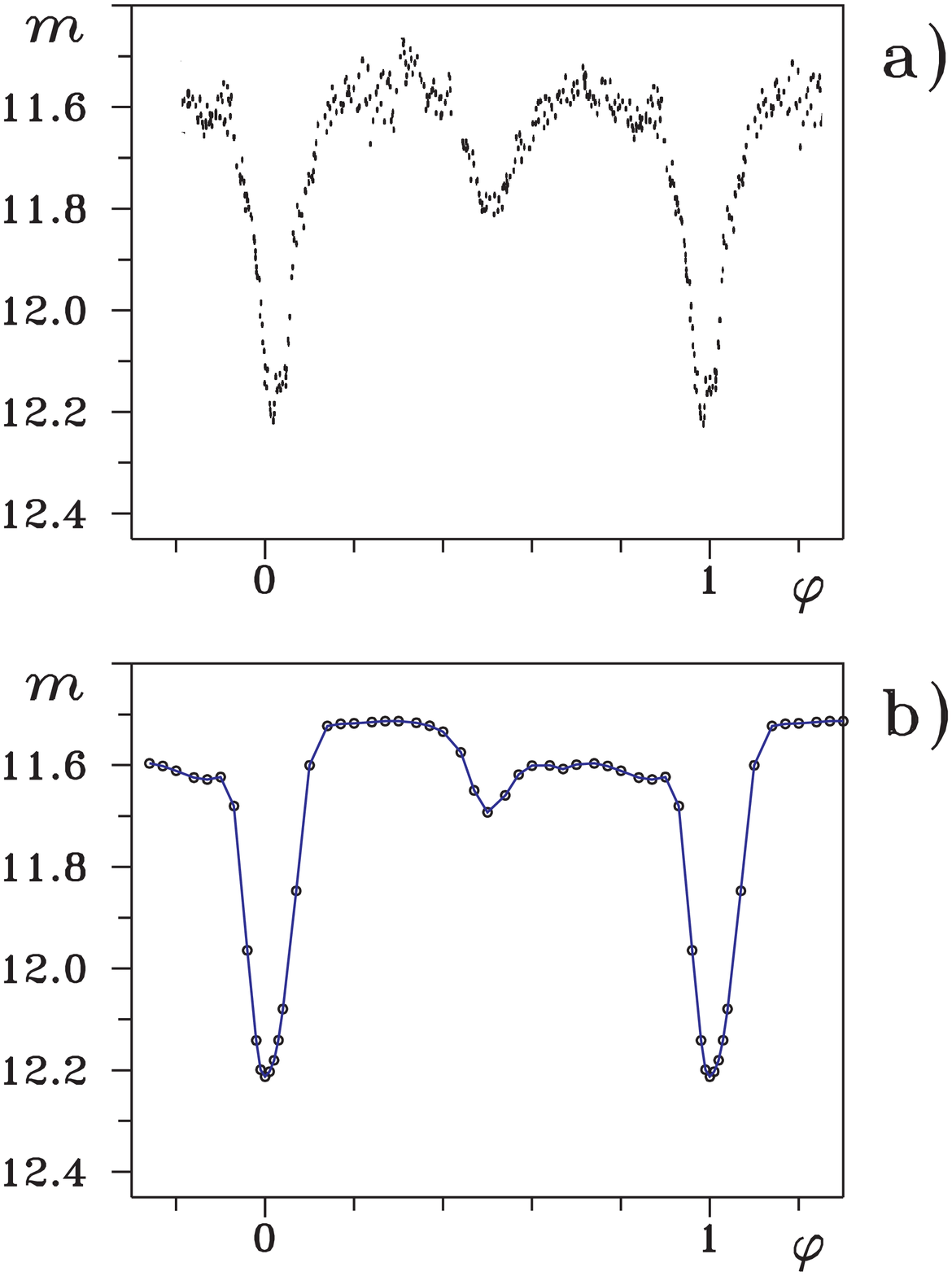}
\caption{Cataclysmic variable with an anomalously placed hump: (a)
observed [34] and (b) theoretical light curves of the dwarf nova RW
Tri in the $K$ filter.}
\end{figure}

\subsubsection{Systems with anomalous hump positions}

   The classical light curves presented above can be
qualitatively rather well described by our model for flow
without a hot spot. Note also that our results are in good
agreement with calculations for standard models with a hot spot,
which were developed especially to interpret such light curves.
Therefore, for our further verification of the adequacy of our
shock-wave flow model, we will consider light curves that could
not be explained in the framework of standard hot-spot models.
These correspond to systems displaying an extended hump at phase
0.5 and systems in which the eclipse begins on the ascending
branch of the hump.

   RW Tri is one of the most remarkable cataclysmic variables in
this context.  This system is a so-called anti-flare star, or UX
UMa type star. A hump is clearly visible in both the optical and
infrared light curves of RW Tri after the egress of the white
dwarf eclipse (Fig.~9a [34], 10a [16]). The main parameters of
this system were derived primarily from the width of the white
dwarf eclipse [7, 16], which is $\Delta\varphi=0.078\pm 0.002$
of the orbital period; $q = 5$ for $i = 90^\circ$, $q = 2$ for
$i = 75^\circ$, and $q = 0.8$ for $i = 70^\circ$.  According to
the studies of Horne and Stiening [16], the radius of the disk
is rather large, and reaches nearly 60
is the distance between the inner Langrange point and the center
of mass of the white dwarf.  The red dwarf has spectral type M0
V, corresponding to a surface temperature of $T_{rd} \sim 3700$~K.

 Figure 9b presents the theoretical light curve in the $K$
filter calculated for parameters close to those determined
earlier for the system: $q=1.2$ for $i=74^\circ$ and
$R_d=0.233a$ ($\sim 0.45\xi$). The remaining model parameters
are given in the Table~1. We assumed that the appearance of the
hump after the egress of the white dwarf eclipse could be due to
strong emission from the side of the bulge turned toward the
incident stream. Therefore, we adopted the model temperature of
the bulge on this side to be $T_{up} \sim 37000$~K, and on the
other side to be about half this, $T_{down}\sim 17000$~K. The
calculated light curve (fig.~9b) is in good agreement with the
observed curve. In particular, we can clearly see the increase
in brightness during the eclipse egress above the brightness
level during the eclipse ingress. Figure 9c presents part of the
theoretical $K$ light curve of RW Tri near the eclipse of the
bulge (in intensity units). The shape of the minimum is
determined primarily by the wide brightness minimum of the red
dwarf, which is associated with the passage into the line of
sight of cold areas on the back hemisphere of the red dwarf that
have not been heated by the white dwarf, and also with the
eclipse of the unevenly heated bulge, on which the narrower
eclipse of the disk is superposed (Fig.~9d).

\figurenum{9c}
\begin{figure}[h]
\plotone{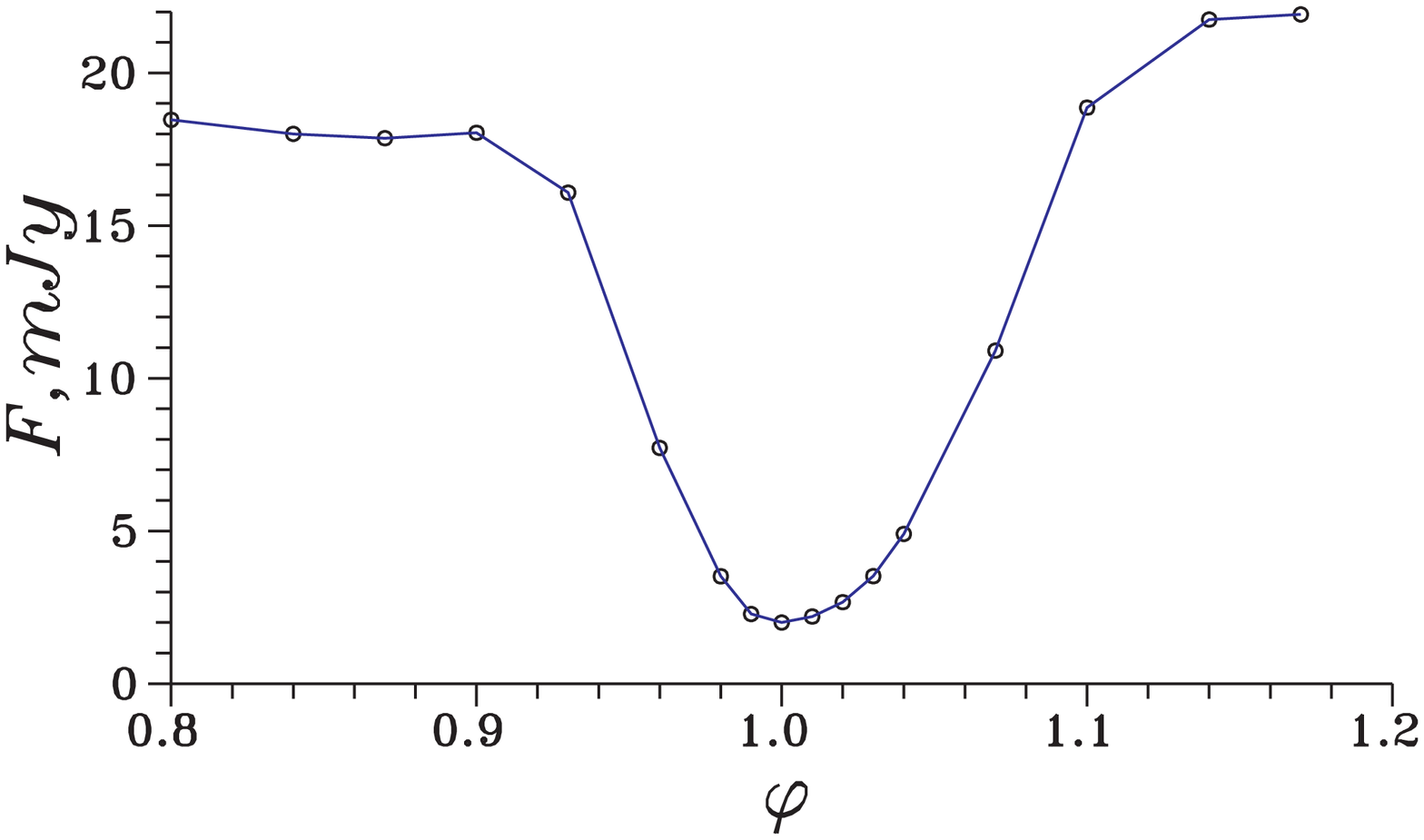}
\caption{Part of the theoretical $K$ light curve for RW Tri in the
region of the bulge eclipse.}
\end{figure}

\figurenum{9d}
\begin{figure}[h]
\plotone{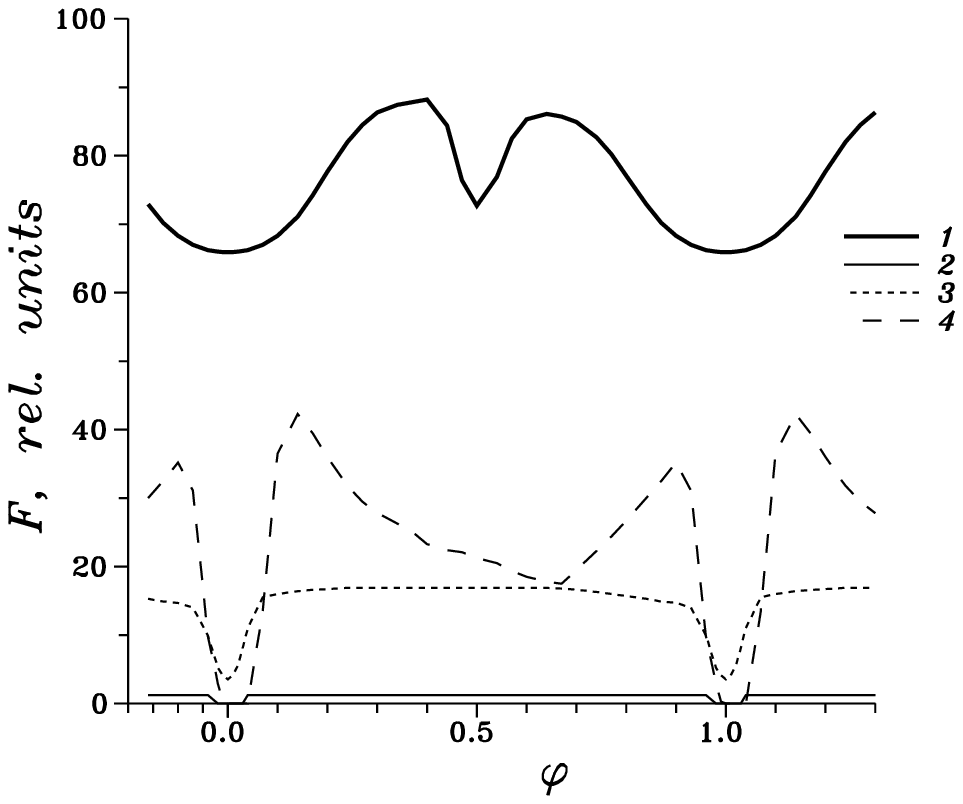}
\caption{Separate dependences of the brightnesses of the red
dwarf (1), white dwarf (2), accretion disk (3), and bulge (4) on
orbital phase for Fig.~9b.}
\end{figure}

 Figure 10 presents the observed [16] and theoretical light
curves for this system (also in intensity units) at optical
wavelengths (the $B$ filter). When constructing the model curve
for the $B$ filter, we used somewhat different parameters for the
bulge: $a_b=0.163a$, $c_b=0.029a$, and $\alpha_b=18^\circ$;
the brightness temperature on the side of the incident stream
was decreased to $T_{up} \sim 23000$~K, while the temperature on
the opposite side remained the same. It is clear that the
qualitative agreement between the model and observed light
curves is maintained at optical wavelengths.

\figurenum{10}
\begin{figure}[h]
\plotone{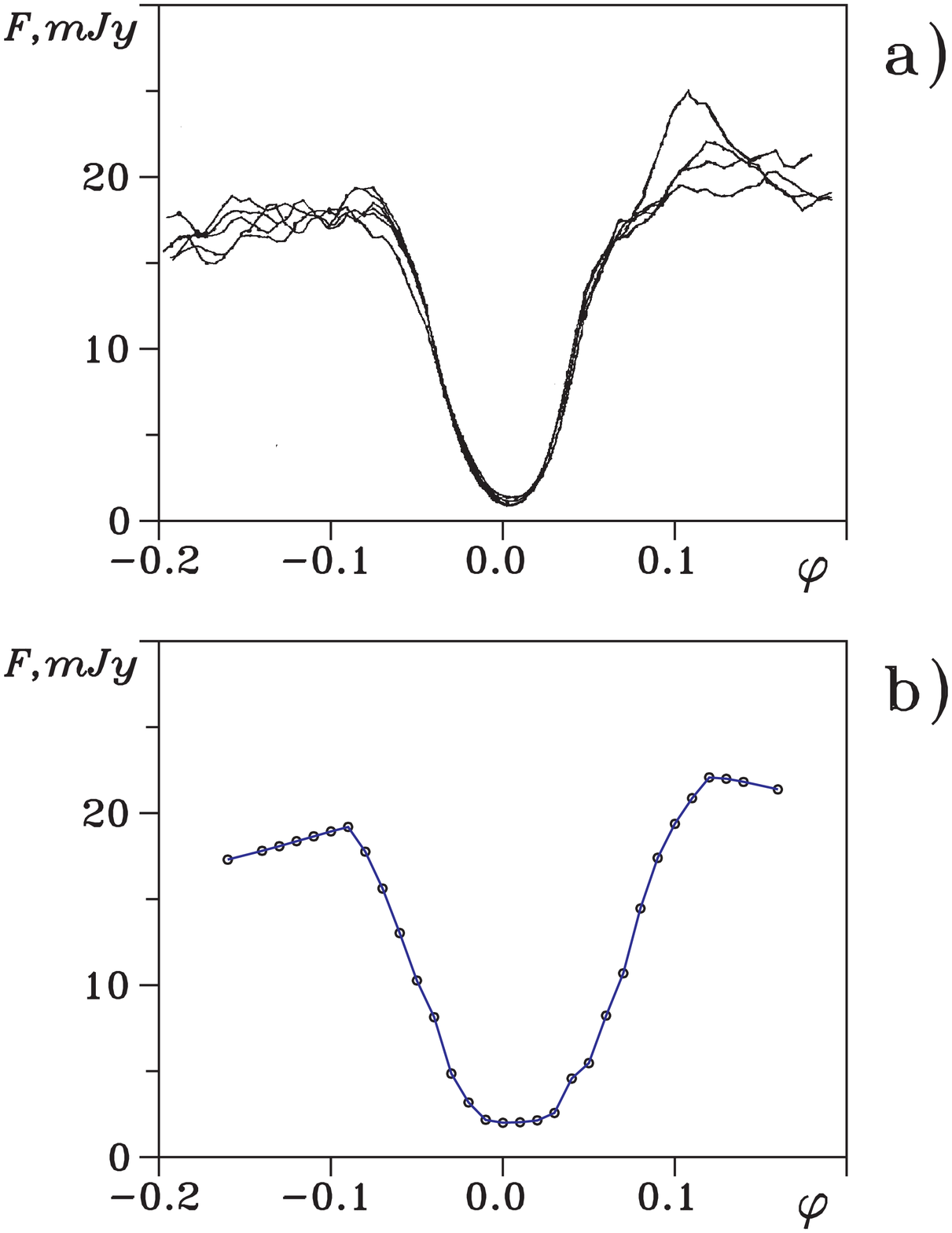}
\caption{Cataclysmic variable with an anomalously placed hump: (a)
observed [16] and (b) theoretical light curves of RW Tri in the $B$
filter in the region of the bulge eclipse.}
\end{figure}

\subsubsection{Systems with an intermediate hump in their light curves}

   Even more interesting from the point of view of testing our
model's ability to describe "irregularities" in the light curves
of cataclysmic variables is the light curve of OY Car, which is
characterized by the episodic appearance of a secondary hump
with lower amplitude near orbital phase $\varphi \sim 0.5$
(Fig.~11a).  In the framework of our model, this can also be
explained by appropriate ratios of the luminosities of the model
components in a system with $i$ close to $90^\circ$.

\figurenum{11}
\begin{figure}[h]
\plotone{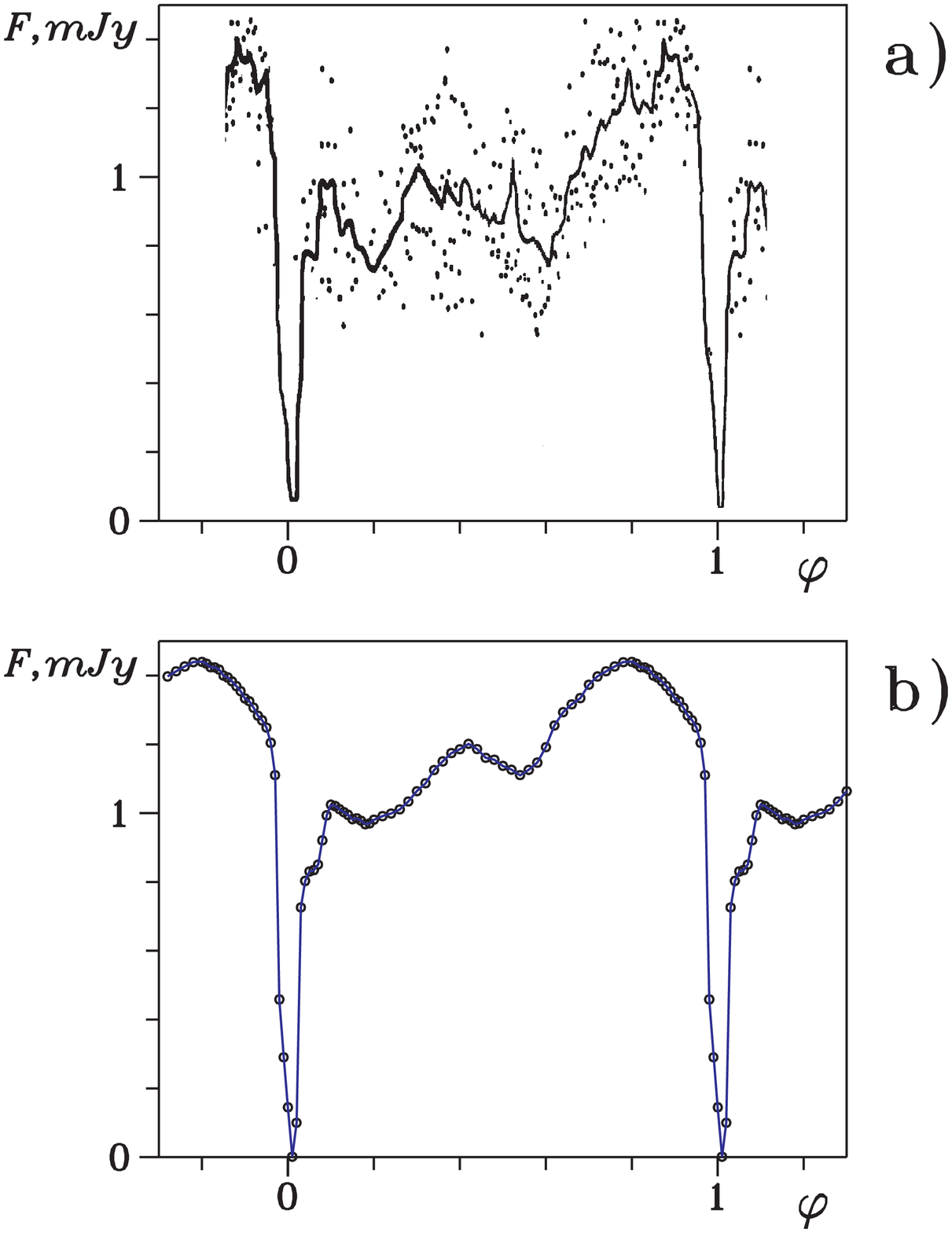}
\caption{System with an intermediate hump in the light curve: (a)
observed [17] and (b) theoretical light curves of the dwarf nova OY
Car in the $V$ filter.}
\end{figure}

   Numerous previous studies of the region of the eclipse of the
white dwarf indicate the parameters for OY Car [7] to be
$q=9.8\pm 0.3$, $i=83^\circ$, Sp M5~V ($T_{rd}\sim 3000$~K), and
$R_d=0.355a$. We fixed the values for $q$, $i$, $T_{rd}$, and
$R_d$ and fitted the remaining parameters in order to
qualitatively reproduce the main features of the light curve of
OY Car (see Table~1). The resulting theoretical light curve is
presented in Fig.~11b, and is clearly in good qualitative
agreement with the observed curve. The secondary hump in the
light curve is associated with heating of the bulge by radiation
from the white dwarf.

\figurenum{11c}
\begin{figure}[h]
\plotone{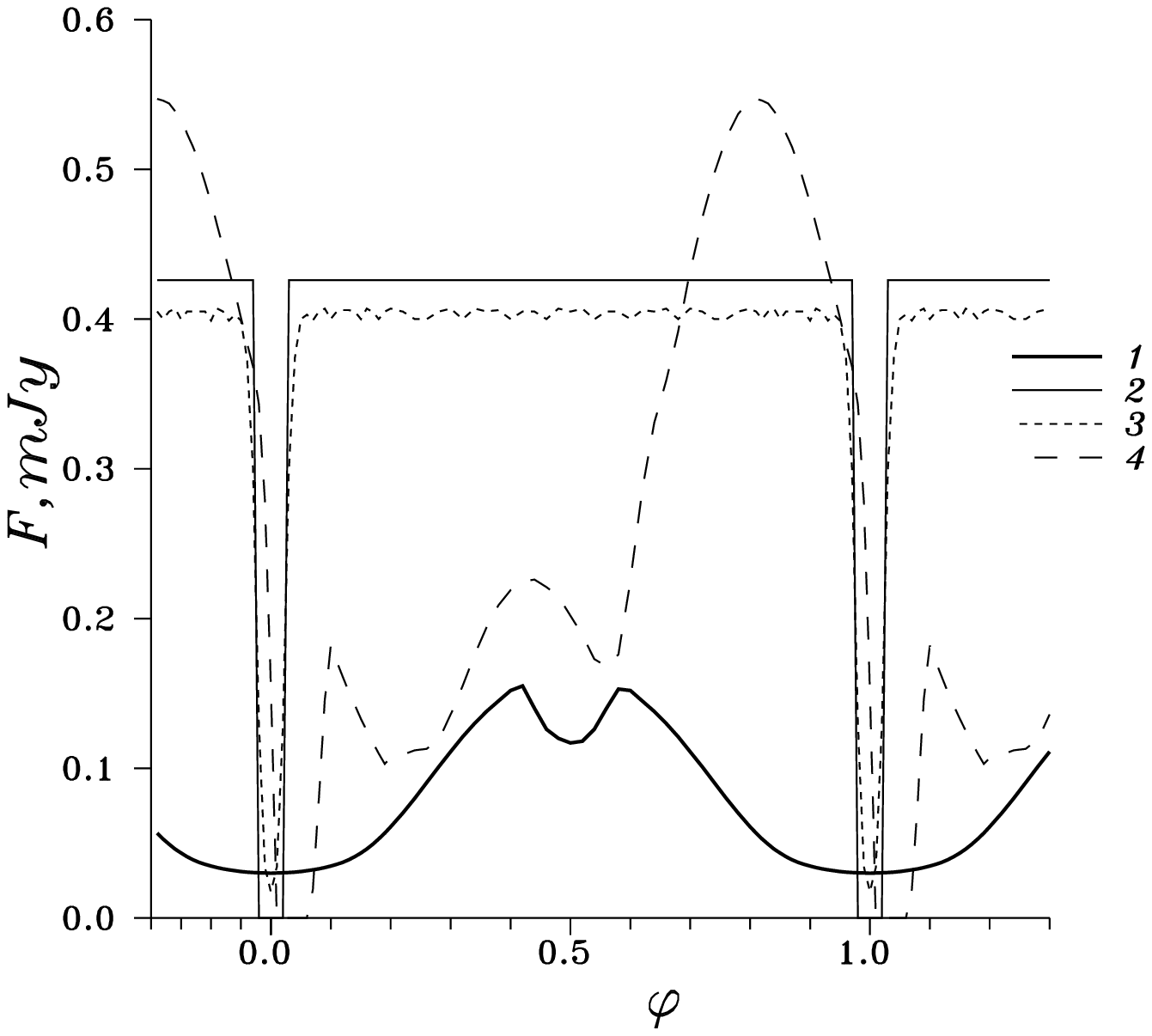}
\caption{Separate dependences of the brightnesses of the red dwarf (1),
white dwarf (2), accretion disk (3), and bulge (4) on orbital phase for
Fig.~11b.}
\end{figure}

   Figure 11c shows the contributions of individual components
of the system to the overall flux at various orbital phases. The
white dwarf makes the largest contribution to the total flux at
virtually all phases, except the phase when it is eclipsed by
the red dwarf and phases when the line of sight intersects the
hottest region of the bulge ($\varphi \sim 0.7-0.9$). The
contribution of the disk outside of eclipse is somewhat smaller.
The flux from the normal star at the maximum is nearly a third
the flux of the white dwarf. Variations in the brightness of the
red dwarf are primarily associated with the reflection effect,
i.e., the heating of the red dwarf by radiation from the hot
white dwarf. The most interesting variations are the
orbital-phase variations of the bulge, associated with its
turning relative to the line of sight. The secondary, smaller,
hump at phase $\varphi \sim 0.4$ corresponds to the situation
when the line of sight intersects the part of the bulge that is
higher than the upper edge of the disk, and is heated by both
radiation from the white dwarf and the shock interaction with
the common envelope gas.

   There is no doubt that our results represent only a crude,
rather schematic attempt to explain the observed features of the
light curves of cataclysmic variables by invoking the presence
of an optically thick formation with substantial geometrical
size outside the accretion disk --- a shock wave at the edge of
the gaseous stream flowing around the disk. However, even in the
framework of our simplified photometric model, we are able to
obtain a variety of light curves that reflect rather fully all
the main features of the brightness variations of cataclysmic
variables.

\section{Conclusion}

   Our three-dimensional numerical simulations of the
gas--dynamical flow of matter in a cataclysmic binary system
similar to Z Cha provide evidence that there is no shock
interaction between the stream of matter flowing from $L_1$ and
the accretion disk in a self-consistent solution for this
system. We first suggested that this was the case based on our
studies of low-mass X-ray binary systems [24, 25], however at
that time, the question of the applicability of our model to
other types of binary systems remained open. Our results here
have confirmed that the solutions for different types of
semidetached binary systems are qualitatively similar,
indicating that the behavior ontained in our simulations is
universal in these systems.

   Summarizing the main properties of the resulting flow
patterns in our model, we first note that the rarified
common-envelope gas significantly influences the structure of
the gas flows in a non-magnetic system in a
steady-state flow regime.  The common-envelope gas interacts
with the stream flowing from the vicinity of $L_1$ and deflects
this stream, so that its interaction with the outer edge of the
accretion disk is shock-free (tangential), and no hot spot is
formed in the disk. At the same time, the interaction of the
common-envelope gas with the stream leads to the formation of
an extended shock wave along the edge of the stream. One of the
characteristic properties of this shock is its variable
intensity along its length, with the main energy release
occurring in a compact region of the shock near the accretion
disk.

   We compared synthetic light curves based on our flow model
with a variety of observed light curves for cataclysmic binary
systems. These comparisons show that our flow model without a
hot spot can describe the light curves rather well. In addition,
we are able to create synthesized light curves that are in good
agreement with the observed curves in cases when this is not
possible in the framework of standard models with an
accretion disk hot spot. We, therefore, conclude that this
observational confirmation of the adequacy of our
self-consistent, gas--dynamical flow model provides a firm
basis to prefer this model over standard hot-spot models in the
investigation of semidetached close binary systems.

\acknowledgments

This work was supported by the Russian Foundation for Basic Research
(project codes 96-02-16140, 96-02-19017)
and by the Program of Support to Leading
Scientific Schools of the Russian Federation
\linebreak (project no. 96-15-96489).

\end{document}